# Optimizing Shanghai's Household Waste Recycling Collection Program by Decision-Making based on Mathematical Modeling

Jiaxuan Chen[1¶*] , Ling Zhou Shen[1&], Jinchen Liu[1&]

[1]Department of Middle School IB-MYP Curriculum, Shanghai World Foreign Language Academy, Shanghai, People's Republic of China

* Corresponding Author

E-mail: chenjiaxuan123456@outlook.com

[¶]This author contributed the most to this work.

[&]These authors contributed equally to this work

# Abstract

In this article, we will discuss the optimization of Shanghai's recycling collection program, with the core of the task as making a decision among the choice of the alternatives. We will be showing a vivid and comprehensive application of the classical mathematical multi-criteria decision model: Analytical Hierarchy Process (AHP), using the eigenvector method. We will also seek the key criteria for the sustainability development of human society, by assessing the important elements of waste recycling.

First, we considered the evaluation for a quantified score of the benefits and costs of recycling household glass wastes in Shanghai, respectively. In the evaluation of each score, we both adopted the AHP method to build a hierarchical structure of the problem we're facing. We first identified the key assessment criteria of the evaluation, on various perspectives including direct money costs and benefits, and further environmental and indirect considerations. Then, we distributed questionnaires to our school science teachers, taking the geometric mean, to build the pairwise comparison matrix of the criterion. A consistency check is done by eigenvector method to ensure that the matrix is consistent for weight calculation. By finding the normalized eigenvector of the matrix, we obtained our weight vector, which is used for the evaluation of the scores. After the theoretical modeling works are done, we began collecting the essential datasets for the evaluation of each score, by doing research on the official statistics, Internet information, market information and news reports. Sometimes, we proceed a logical pre-procession of the data from other data, if the data wanted isn't directly accessible.

Then, we crucially considered the generalization of our mathematical model. We considered from several perspectives, including the extension of assessment criteria, and the consideration of the dynamic interdependency between the wastes, inside a limited transportation container. After using AHP again, we finish with new weight vectors. A crucial and logical data collection process is done, with a min-max normalization. Now, the data is ready for scoring by adding weights. A benefit and a cost score are both evaluated for each type of data, with the theoretical maximum value of 1. By comparing the data's final score, which is the difference between the benefit and difference, we are eligible to make our final decisions. Our model can also be adapted to optimize the waste recycling program for other cities.

# 1. Introduction

## *1.1. Background*

Recycling is defined as the recovery and reprocessing of wastes[1]. It is very important for the sustainable development of human civilization, as the natural resources are very limited on earth, some are fossil resources that's never renewable, and even the renewable resources can be damaged seriously when over-used by humanity. Recycling wastes not only conserves the resources as a direct effect, indirectly, it can also save energy, reduce pollution and save land resources[2][3].

## *1.2. Social Issue Restatement*

In the big cities of China, the population is dense, and the trash collection work is very limited. In order to achieve an optimized sustainability with limited resources, we have to make a choice among the recyclable wastes. The city of our choice is Shanghai, for its high significance in the development of economy, with a representative effect of the whole situation of China. In this essay, we will first warmup with the case of glass wastes recycling, including bottles and jars, trying to quantify the benefits and costs into a numerical score, not only on the aspect of money, but also including further considerations on environment pollutions, and land resources. Then, we will develop a more general model, to evaluate the optimization of each type of waste, including further considerations of transportation, benefits and costs, also the co-pollution and contamination costs to be transported in the same vehicle, and finally make the decision to recycle the most beneficial ones of them.

# 2. Assumptions and Justifications

**Assumption 1: The constant depth of waste landfill:** We assume that the ground of the waste landfilling area is flat, with a constant depth everywhere. Although geographically the ground is never perfectly flat for a equal-depth landfill, due to the randomness of location of landfilling and to simplify the model, we assume that the depth is constant. This is quite reasonable, because the probabilities to witness different depths are in normal distribution, and for number of samples reaching the infinite, according to the law of large numbers, the expectation is the average value of all the depths. With a constant depth, the land resources taken by a type of trash can be directly proportionally represented the inverse of its density, which we will discuss later in our model.

**Assumption 2: The stability of the society market state:** We assume that the current market state can be approximately represented by the past data we find, and is relatively stable, without very significant changes. Our assessment system highly depends on the input data reflecting the Chinese society's market, and this is quite a dynamic factor that we can't take into our model except by using very complex probability distribution methods. The stability of the market ensures the validity of the input of our model, which is a factor we can't control without such an assumption.

**Assumption 3: Artificial Selection and Contamination of waste:** Assume that the selection and contamination process of waste are all done in artificial for by workers. Although there are automatic machines to contaminate the wastes, the most widely applied method in China is still artificial selection. These 2 methods will lead to the consideration of different criteria to assess the time cost of selection for different types of wastes. In order to confirm the universal validity of the criterion in all cases, we assume artificial selection to simplify our model.

**Assumption 4: Equal Non-Organic Cleanliness of Wastes:** Assume there's no organic material in the wastes we collected here for recycling in Shanghai, all organic wastes are pre-separated before collection. This ensures that there will be no unequal pollution depending on the property and daily life use of each trash, so that we don't have to consider the factor of co-pollution as it's fair to all alternatives.

**Assumption 5: Burning of Paper not Allowed:** Assume the treatment of burning for papers isn't used in Shanghai. This method isn't widely used in China anymore, for its high air pollution.

**Assumption 6: The sharing of resources and market state throughout China:** Assume the market state and resources usage situations are the same in the whole country of China, with all the cities sharing the mineral resources including Shanghai. This ensures the data we took specifically for the whole China is also valid to represent the situation in Shanghai.

# 3. Definition of Variables

As shown in Table 1 are the definitions of the variables to be used throughout the models. Temporary or less-important variables may be defined later in the corresponding part of the essay.

***Table 1 Variable Definitions***

| Variables | Descriptions |
|---|---|
| $P_i$ | The average price of the recycled waste material i(CN¥/ton) |
| $U_i$ | The usage demand of the recycled waste material i (kiloton/year) |
| $I_i$ | The limitedness index of natural resources in production of waste i |
| $E_i$ | The energy use during the primary production of the material i(GJ/ton) |
| $R_i$ | The recycling rate of the waste i(%) |
| $T_i$ | The $CO_2$eq emission by regular waste treatment of i(kg $CO_2$eq/ton) |
| $Mt_i$ | The $CO_2$-eq emission of production of the material i(kg $CO_2$eq/ton) |
| $C_i$ | The money cost of landfill/burning of waste i(CN¥/ton) |
| $L_i$ | The land resources taken by landfill of waste i ($cm^3$/kg) |
| $D_i$ | The effect duration of the landfill of waste i(year) |
| $S_i$ | The reciprocal of average volume of a single waste I (1/$cm^3$) |
| $Eq_i$ | The cost of recycling mechanical equipment for waste i (CN¥/equipment) |
| $Er_i$ | The electricity used during recycling process of each ton of waste (kWh/ton) |
| $Cr_i$ | The $CO_2$eq emission of recycling of each ton of waste i (kg $CO_2$eq/ton) |
| $Ar_i$ | The acidic gases emission of recycling of each ton of waste i(kg $N_xO_x$ & $SO_2$ /ton) |
| $At_i$ | The acidic & other poisonous gases emission of regular waste i's treatment (kg/ton) |
| N | Number of questionnaire samples collected |
| $Q_i$ | The $i^{th}$ questionnaire sample matrix |
| M | The pairwise comparison matrix |
| n | The dimension of matrix M |
| $V_i$ | The eigenvectors of matrix M (V is a matrix whose columns are eigenvectors) |
| $\lambda_i$ | The eigenvalues of matrix M |
| $\lambda_{max}$ | The principal eigenvalue of matrix M |
| $RI_i$ | The random consistency index of matrix M with dimension i |
| CI | The consistency index of matrix M |
| CR | The consistency ratio of matrix M |
| $W_i$ | The weight of the $i^{th}$ criterion |
| $BS_i$ | The recycling benefit evaluation score of the waste type i |
| $CS_i$ | The recycling cost evaluation score of waste type i |
| $Cri_{i,j}$ | The data value for criterion j on waste type i |
| $NormzalizedCri_i$ | The normalized score for criterion i |
| $Cmp_{i,j}$ | The scarity index of composition material j of waste i |
| $Wt_{i,j}$ | The proportion weight of material j in waste I (%) |
| $FS_i$ | The final evaluation score of waste i |

# 4. Specific Case Interpretation: Glass Recycling Evaluation Model

In this section, we're going to evaluate quantified scores of the benefits and costs of household glass waste recycling in Shanghai, including glass bottles and jars. We will discuss the 2 aspects respectively, with a benefit score BS and a cost score CS as final results. In each model, we will use the Analytical Hierarchy Process (AHP) model for the weight calculation, as it's a multi-criteria decision-making model that can decompose the complex overall evaluation into organized sub-steps, with a hierarchical structure[5][78][79]. In each model, we will first identify the key assessment criteria that will take place in the process, and then we will use a pairwise comparison matrix to compare the importance of the criteria, using the 17-point scale (Table 2). This comparison method converts subjective comparison to more crucial weights, which can help us to reach out a more objective and accurate score. After that, we will use the eigenvector method to calculate the final criterion weights. Then, we will collect the data and score for each criterion, with some necessary pre-processions. At last, we will demonstrate the process to add weight to the scores to obtain the final overall scores, but as there's no opportunity to normalize the data by only 1 alternative, we will come up with the normalized dataset and calculate the final scores in the section 5, where we have multiple datasets as evidence of normalization.

***Table 2 19-point Scale Level Description[5][79]***

| Intensity of Importance | Definition |
|---|---|
| 1 | Equal Importance |
| 3 | Weak Importance of one over another |
| 5 | Essential or strong importance |
| 7 | Very strong or demonstrated importance |
| 9 | Absolute importance |
| 2, 4, 6, 8 | Intermediate values between adjacent scale values |
| Reciprocals of above nonzero | If I has one of the above nonzero numbers assigned to it when compared with j, then j has the reciprocal value when compared with i |

## 4.1. Benefit Model of Glass Recycling

### 4.1.1. Definition of Glass Recycling Benefits

Before we process the development of the model, to specialize the objective of this model, for its accuracy and specificality of, we have to clear the concept of "Glass Recycling Benefits". We define "Glass Recycling Benefits" as an overall evaluation of direct and indirect benefits in aspects of reuse, environment and land resources, obtained by the action of recycling household glass wastes, including bottles and jars that's made of glass.

## 4.1.2. Identification of Assessment Criteria

The determination of assessment criteria is key to the decision process, supporting the base of the hierarchical structured logic. To construct the hierarchical structure of the model, we identify our key assessment criteria for the model, classified in the following 3 aspects of consideration:

### 1. Reuse Benefits

By the reuse of the glass products after recycling, we obtain several direct benefits. These are the criteria to be considered in our benefit modeling.

• **The price of recycled material (P):** This variable represents the social economic benefits of recycling the household glass wastes. The re-produced glass products can bring their value to the society, and help the development of economy, while also providing direct profit. These aspects can all be shown through the price of the recycled products, measured in CN¥/ton.

• **The usage demand of recycled material (D):** This also shows the economic benefits of recycling household glass wastes from another perspective. The demand of recycled material can reflect the potential economic value that can be obtained by re-selling them. It can be represented by the number of products sold, measured in ton/year.

• **The limitedness of the natural resource (I):** Natural resources are also present in the primary production of wastes. Some natural resources can be highly rare, and recycling would be essential for their limitedness, on the aspect of a sustainable development of the region's natural resources. We consider this variable on the background of the whole country, because the cities in China share the natural resources together. This criterion will have an input of a limitedness index, the evaluation method of which will be discussed later.

• **The energy use during the primary production of the material (E):** Producing glass would cost some energy. By reusing the glass wastes, we save this portion of energy, which is considered as a benefit of recycling. The energy will be measured in GJ per ton of glass.

• **The recycling rate of the waste (R)**: During the recycling process, there could be a partial wastage of the material, which is a regular phenomenon. The final amount of recycled material is only a part of the original material, with part of that wasted. So, the recycling rate of glass waste is also an important factor to consider in the benefits.

### 2. Avoidance of Environmental Damage

By recycling household glass wastes, we avoided the environmental damage that's present in the traditional glass waste treatment of landfilling (Burning treatment isn't applicable to glass as it's a non-flammable material). This avoidance of environmental damage can also be considered as benefits of glass waste recycling.

• **The $CO_2$-eq (Carbon Dioxide Equivalent) emission of regular waste treatment (T):** The regular waste treatment of glass is landfilling, so the related actions do have some carbon emissions, which are avoided by the alternative action of recycling. The effect on the environment of the greenhouse gases (GHGs) released during the process can be measured by carbon dioxide equivalent ($CO_2$-eq), which is a value obtained by adding weights to the amount of different GHGs released, based on their effects. This variable is measured by kg $CO_2$-eq released per kg glass wastes.

• **The $CO_2$-eq (Carbon Dioxide Equivalent) emission of production of the material (Mt):** Apart from the landfilling, the primary production of glass also produces GHGs, and the emission of them can also be avoided by recycling glass wastes. This variable is also measured by kg $CO_2$-eq released per kg glass.

### 3. Avoidance of Landfill Costs

• **The money cost of landfill (C):** Landfill would cost some money, and this cost is saved by choosing to recycle as an alternative. This benefit is measured by CN¥ per ton of glass wastes.

• **The land resources taken by landfilling (L):** Another significant disadvantage of landfill is that it takes up a lot of space, which wastes the precious land resources. We will consider this factor as related to the inversed density of the trash, which is the volume it takes up in the space per unit mass ($cm^3$/kg). This can proportionally reflect the land resources it uses, as we've assumed the depth of waste landfilling is the same.

• **The effect duration of the landfill (D):** When landfilling is avoided, the benefit also includes the avoidance of a possibly long-term or nearly permanent damage to the land resources. To include this point of view in our model, we add this variable as the duration for the trash to naturally decompose in the earth, which is the period in which the land is affected. This variable is measured in years, as it's usually a long time.

## 4.1.3. Constructing the Analytic Hierarchy Process Model

### 1. Structuring the Hierarchy

Using the assessment criteria we identified, we can now construct the hierarchical structure used in our model (Fig 1)[5]. As here we use AHP model to evaluate only 1 alternative, the 3rd level of hierarchy (alternatives) isn't essential.

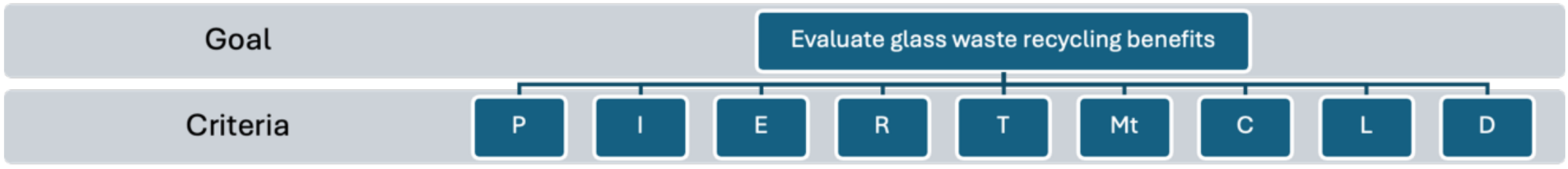

*Fig 1 Hierarchical Structuring of Glass Waste Recycling Benefit Evaluation Model*

### 2. Pairwise Comparison of Criteria

We will use the 19-point scale to build the pairwise comparison matrix. We designed a questionnaire to collect the opinion of professional respondents including our school teachers in Shanghai. We collected 4 samples (shown in Appendix 1.1). We take the geometric mean of each element among the matrices (here N=4)[6]:

$$m_{i,j} = \sqrt[N]{\prod_{k=1}^{N} q_{k_{i,j}}} = \sqrt[4]{\prod_{k=1}^{4} q_{k_{i,j}}} \ (i,j \in [P,U,I,E,R,T,Mt,C,L,D])(4.1)$$

After performing that calculation using a Microsoft Office Excel spreadsheet, we have our final $10 \times 10$ pairwise comparison matrix in 19-point scale, rounded to 4 decimal places, where $m_{i,j}$ represent the comparison of criterion i to criterion j:

$$M = \begin{bmatrix} m_{P,P} & m_{U,P} & m_{I,P} & m_{E,P} & m_{R,P} & m_{T,P} & m_{Mt,P} & m_{C,P} & m_{L,P} & m_{D,P} \\ m_{P,U} & m_{U,U} & m_{I,U} & m_{E,U} & m_{R,U} & m_{T,U} & m_{Mt,U} & m_{C,U} & m_{L,U} & m_{D,U} \\ m_{P,I} & m_{U,I} & m_{I,I} & m_{E,I} & m_{R,I} & m_{T,I} & m_{Mt,I} & m_{C,I} & m_{L,I} & m_{D,I} \\ m_{P,E} & m_{U,E} & m_{I,E} & m_{E,E} & m_{R,E} & m_{T,E} & m_{Mt,E} & m_{C,E} & m_{L,E} & m_{D,E} \\ m_{P,R} & m_{U,R} & m_{I,R} & m_{E,R} & m_{R,R} & m_{T,R} & m_{Mt,R} & m_{C,R} & m_{L,R} & m_{D,R} \\ m_{P,T} & m_{U,T} & m_{I,T} & m_{E,T} & m_{R,T} & m_{T,T} & m_{Mt,T} & m_{C,T} & m_{L,T} & m_{D,T} \\ m_{P,Mt} & m_{U,Mt} & m_{I,Mt} & m_{E,Mt} & m_{R,Mt} & m_{T,Mt} & m_{Mt,Mt} & m_{C,Mt} & m_{L,Mt} & m_{D,Mt} \\ m_{P,C} & m_{U,C} & m_{I,C} & m_{E,C} & m_{R,C} & m_{T,C} & m_{Mt,C} & m_{C,C} & m_{L,C} & m_{D,C} \\ m_{P,L} & m_{U,L} & m_{I,L} & m_{E,L} & m_{R,L} & m_{T,L} & m_{Mt,L} & m_{C,L} & m_{L,L} & m_{D,L} \\ m_{P,D} & m_{U,D} & m_{I,D} & m_{E,D} & m_{R,D} & m_{T,D} & m_{Mt,D} & m_{C,D} & m_{L,D} & m_{D,D} \end{bmatrix}$$

$$= \begin{bmatrix} 1.0000 & 0.7071 & 2.7832 & 2.1407 & 0.8409 & 3.9843 & 2.2795 & 0.5774 & 1.6226 & 3.1179 \\ 1.4142 & 1.0000 & 1.5651 & 1.8612 & 1.1067 & 1.5137 & 1.3161 & 0.5373 & 2.7108 & 1.8612 \\ 0.3593 & 0.6389 & 1.0000 & 0.5466 & 0.5411 & 1.5651 & 2.3784 & 0.5318 & 0.9306 & 1.9343 \\ 0.4671 & 0.5373 & 1.8294 & 1.0000 & 1.5651 & 1.7321 & 1.4142 & 0.5000 & 1.1892 & 2.4746 \\ 1.1892 & 0.9036 & 1.8481 & 0.6389 & 1.0000 & 1.5651 & 1.2574 & 0.6606 & 1.0393 & 2.3784 \\ 0.2510 & 0.6606 & 0.6389 & 0.5774 & 0.6389 & 1.0000 & 1.0000 & 0.4518 & 0.9036 & 1.2574 \\ 0.4387 & 0.7598 & 0.4204 & 0.7071 & 0.7953 & 1.0000 & 1.0000 & 0.3593 & 0.8409 & 1.7321 \\ 1.7321 & 1.8612 & 1.8803 & 2.0000 & 1.5137 & 2.2134 & 2.7832 & 1.0000 & 2.7108 & 2.7108 \\ 0.6148 & 0.3689 & 1.0746 & 0.8409 & 0.9622 & 1.1067 & 1.1892 & 0.3689 & 1.0000 & 2.1407 \\ 0.3207 & 0.5373 & 0.5170 & 0.4041 & 0.4204 & 0.7953 & 0.5774 & 0.3689 & 0.4671 & 1.0000 \end{bmatrix} (4.2)$$

The matrix theoretically satisfies the following diagonally symmetrical property[5], although there are flaws after the approximation:

$$m_{i,j} = \frac{1}{m_{j,i}}, and\ m_{i,i} = 1\ for\ all\ i,j \in [P,U,I,E,R,T,Mt,C,L,D]\ (4.3)$$

After we obtained the pairwise comparison matrix, we perform the following steps to obtain our final result. The computation of the following procedures is done by the MATLAB code in Appendix 2.1, using input 1(Remarks: all calculations in the later parts of the essay are by default rounded to 4 decimal places, as this is the default mode of MATLAB calculations). For such a matrix that doesn't satisfy $a_{ik} \cdot a_{kj} = a_{ij}$ for all i, j and k, we should use the eigenvector method to calculate the weights[7]. But before calculating the weights, we first have to confirm the consistency of the matrix by performing a consistency check, as the following.

## 3. Consistency Test

First, we have to perform a consistency check of the matrix to ensure the accuracy and reasonability of our model. The algorithm of the program is as the following[5][6]:

### Step 1: Compute the eigenvalue and eigenvector

These 2 values are possible solutions to the equation:

$$Mv_i = \lambda_i v_i (4.4)$$

Where $V_i$ is the eigenvectors of M, and $\lambda_i$ is the eigenvalue of M[8]. V is a matrix whose columns are eigenvectors of M, and $\lambda$ is a matrix with the eigenvalues distributed on the main diagonal, with the same column index to their corresponding eigenvectors. We define the maximum value of $\lambda_i$ as the principal eigenvalue $\lambda_{max}$. By the computations of MATLAB, we obtain:

$$\lambda_{max} = 10.3774 (4.5)$$

**Step 2: Compute the Consistency Index**

Then we can compute the Consistency Index (CI) of the matrix by applying the formula[5][6], by substituting n=10 and equation (4.4)(n denote the dimension of matrix M):

$$CI = \frac{\lambda_{max} - n}{n - 1} = 0.0419 \ (4.6)$$

**Step 3: Compute the Consistency Ratio**

The Consistency Ratio (CR) is defined as the ratio of Consistency Index (CI) and Random Consistency Index (RI). For here a 10-order matrix, according to Table 3, $RI_{10}$=1.49. Then, substituting equation (4.5), we can compute CR[5][6]:

$$CR = \frac{CI}{RI_n} = \frac{CI}{1.49} = 0.0281 \ (4.7)$$

Here CR=0.0284<0.1, so the matrix in proved as consistent.[5]

***Table 3 Random Consistency Indices***

| Dimension of matrix | Random Consistency Index (RI) |
|---|---|
| 1 | 0.00 |
| 2 | 0.00 |
| 3 | 0.58 |
| 4 | 0.90 |
| 5 | 1.12 |
| 6 | 1.24 |
| 7 | 1.32 |
| 8 | 1.41 |
| 9 | 1.45 |
| 10 | 1.49 |
| 11 | 1.51 |
| 12 | 1.48 |
| 13 | 1.56 |
| 14 | 1.57 |
| 15 | 1.59 |

## 4. Weight calculation

We continue using the eigenvector method to calculate the weight. We consider the weight vector $W_i$ of the assessment criteria as the normalized principal eigenvector of the matrix[5]. So, our algorithm first iterates through all the possible eigenvectors and find the one that corresponds with the maximum eigenvalue, and get its column index:

$$if\ \lambda_x = \lambda_{max}, then\ the\ principal\ eigenvector\ is\ V_{:,x}\ (4.8)$$

Then, we have to perform the normalization of that principal eigenvector with the column index of V as computed in (4.7), as weights always satisfy the property $0 \le W_i \le 1$, which is achieved by the process of normalization[5]:

$$w_j = \frac{v_{j,x}}{\sum_{k=0}^{n} v_{k,x}}\ for\ all\ variable\ j\ \in [1, n]\ and\ constant\ x\ (4.9)$$

Finally, applying formula (4.8), we have our normalized principal eigenvector, which derives our weights of criteria as in Equation (4.10), with weight distribution visualized by Fig 2.

$$W = [0.1491, 0.1300, 0.0813, 0.1016, 0.1066, 0.0617, 0.0668, 0.1788, 0.0776, 0.0464]^T\ (4.10)$$

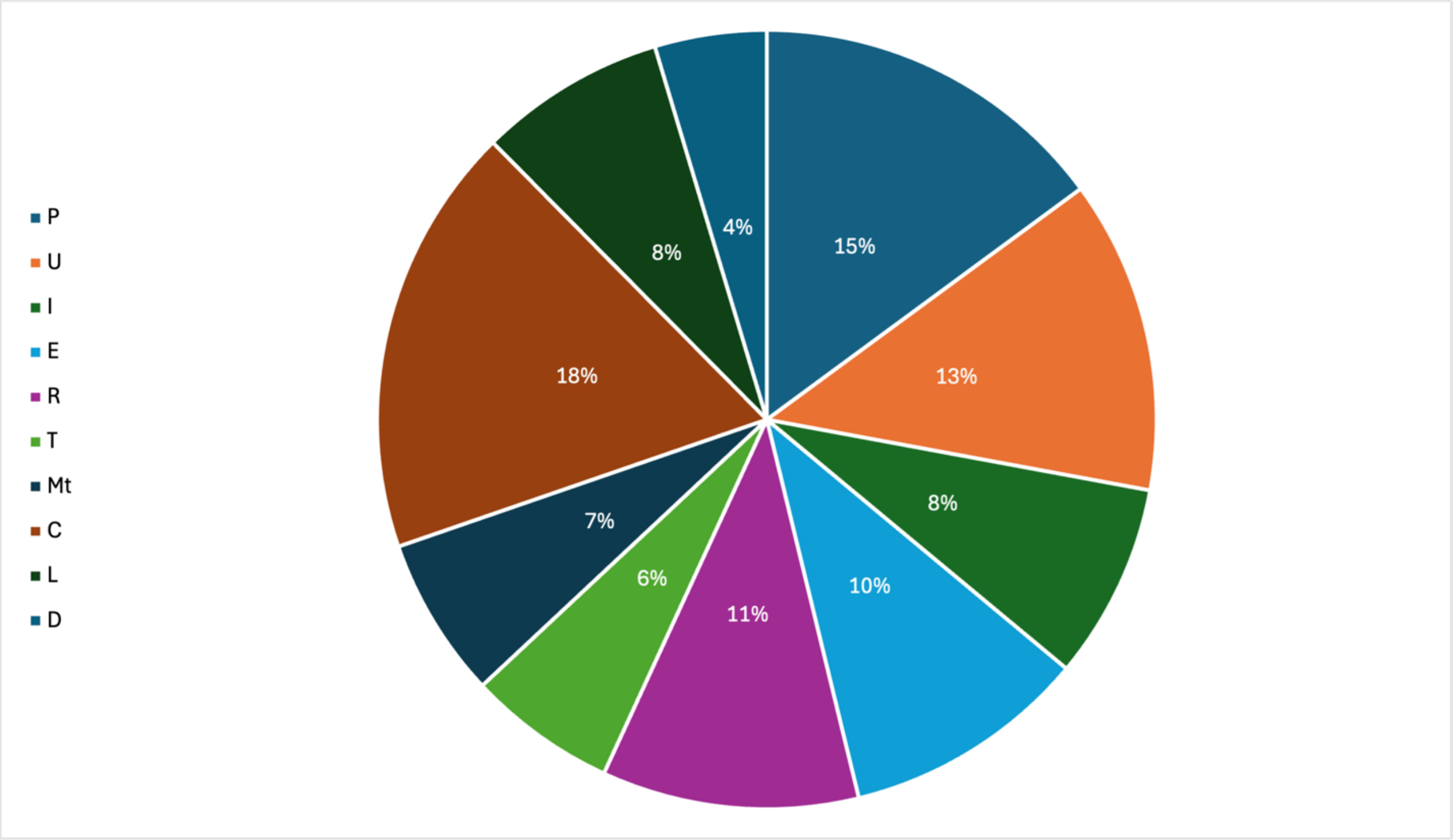

***Fig 2 Weight Distribution of Glass Recycling Benefit Model***

## 4.1.4 Criterion-wise Scoring

We define the numerical scores of the evaluation of each criterion as:

$$Cri_{glass} = [P_{glass}, U_{glass}, I_{glass}, E_{glass}, R_{glass}, T_{glass}, Mt_{glass}, C_{glass}, L_{glass}, D_{glass}]\ (4.9)$$

We will collect these data specifically for China, as the resources and market situation is shared among the whole country, and can represent the demand of Shanghai. By searching on the Internet, by either official statistics or market research, we can directly or indirectly obtain the data we need.

## 1. Directly Measured Data

**P:** In year 2022, the price of glass raw material is 1782.3 CN¥/ton.[9]

**U:** In year 2023, by retrieving from official document, China has demand of 31463.9 kilotons of glass products[10].

**E:** The average Specific Energy Consumption (SEC) is around 7.2 GJ/t for container glass.[11]

**R:** Glass can be recycled endlessly with no loss in quality or purity.[4][12] So, we can consider the recycling rate R=100%, with no material loss during the process.

**T:** Glass wastes are usually treated by landfilling, as the landfilling process of non-organic wastes such as glass has 0 carbon emission[13].

**Mt:** In Chinese industry, producing 1 ton of glass causes emission of 1870 $CO_2$eq. [14]

**C:** The cost of landfilling of wastes in China (No matter of type) is averagely 250 CN¥ (200-300CN¥) per ton. [15]

**L:** The density of glass is 2.7 g/ $cm^3$ at 20°C[16]. The land resources taken by glass wastes is approximately the reciprocal of its density: 0.37 $cm^3$/g = 370 $cm^3$/kg.

**D:** The natural decomposition period of glass is about 4000 years. [17]

## 2. Data with Pre-processions

**I:** The limitedness index should be considered on the perspective of the primitive material composition of glass wastes. For a waste made of different original materials, we define the limited index I as the weighted average of the limitedness of its composition elements. The limitedness is defined as the $Cmp_{glass}$, the scarcity index of the composition material. So, the formula to calculate limitedness index I of waste with n composition materials, each with weight $Wt_{glass,i}$ is:

$$I_{glass} = \frac{\sum_{i=0}^{n} Cmp_{glass,i} \cdot Wt_{glass,i}}{n} \quad (4.10)$$

The scarcity index is calculated by the formula:

$$SI = \frac{D}{R} \text{ where } D \text{ is the demand } (kiloton/year), \text{ and } R \text{ is the known storage } (kiloton)$$

In this case, consider that glass is mainly made of sand (silicon dioxide, $SiO_2$), limestone (calcium carbonate, $CaCO_3$), and sodium carbonate ($Na_2CO_3$), each with proportion weights 75%, 10% and 15%[18]. In year 2022, China produced 97890 kilotons of silicon dioxide sand[19]. There are over 14,000,000 kilotons of silicon dioxide storage in China[20]. In 2023, China has demand of 28,720,000 kilotons of limestone[21]. The known storage of limestone is 504,000,000 kilotons in China[22]. In 2022, China's demand for sodium carbonate is 27262 kiloton [23]. China has 260,000 kiloton of known sodium carbonate storage[24]. Then, we can calculate the scarcity index of these materials in China, rounded to 4 decimal places:

$$Cmp_{glass} = [0.0070, 0.0570, 0.1049] (4.11)$$

Adding weights and taking arithmetic meaning yields:

$$I_{glass} = \frac{0.0070 \cdot 75\% + 0.0570 \cdot 10\% + 0.1049 \cdot 15\%}{3} = 0.0089 \quad (4.12)$$

After all, we have a complete scoring vector of criteria:

$$Cri_{glass} = [1782.3, 31463.9, 0.0089, 7.2, 1, 0, 1870, 250, 370, 4000] (4.13)$$

### 3. Calculating the Score

However, it's significant that the data here isn't normalized, which is not fair for the weighting. The normalization of glass data will be done later together with other types of wastes. The model here is just an implementation of the data collection and pre-procession process, but a crucial result will be produced after normalization. After normalization of the data in $Cri_{glass}$, we can compute the benefit score, which is the output result of our glass waste recycling benefit evaluation model:

$$BS_{glass} = \sum_{i=1}^{10} NormalizedCri_{glass,i} \cdot W_i \ (4.14)$$

# 4.2. Cost Model of Glass Re-Production Process

## 4.2.1 Definition of Glass Re-Production Process Costs

To specialize our model, we clear the concept of "Glass Re-Production Process Costs" before we proceed to the model construction. We define this concept as an overall consideration of economic and environmental costs that take place in the process of re-producing the recycled household glass wastes for reuse.

## 4.2.2. Identification of Assessment Criteria

There are several assessment criteria to be considered. In order to improve the clear system of logic, it's important to classify the criteria into groups. We identify the key assessment criteria as in the following 2 aspects of consideration:

### 1. Economic Costs

• **The time cost of selection and contamination for recycling (S):** Different waste corresponds to a different difficulty of selection and classification for the process of recycling, according to their property. This can be also counted as a cost of recycling. It can be measured as the reciprocal to the volume of each single waste in $cm^3$, as a smaller trash takes a higher time cost, which is a inverse proportional relationship.

• **The cost of recycling mechanical equipment (Eq):** The mechanical equipment used for different types of waste is also different. The equipment cost is also an important portion of recycling cost. It will be recorded in unit of CN¥ per equipment.

• **The cost of electricity used during recycling process (Er):** The electricity cost can be measured in kWh, which is also a major cost included in the process of recycling.

### 2. Environmental Costs

• **The $CO_2$eq emission of recycling (Cr):** During the recycling process, because of the use of electricity and the gas produced by industry, greenhouse gases are released. This is also a cost of recycling, on the environmental aspect.

• **The acidic gases emission of recycling (Ar):** the air pollution of the recycling process also involves the emission of acidic gases, including sulphur dioxide ($SO_2$) and nitrogen oxide ($N_xO_x$).

## 4.2.3. Constructing the Analytic Hierarchy Model

### 1. Structuring the Hierarchy

Using the key assessment criteria we identified, we can build the hierarchical structure of AHP model, as shown in Fig 3, while omitting the third layer of alternatives as it's not essential for only 1 alternative to be considered in this case.

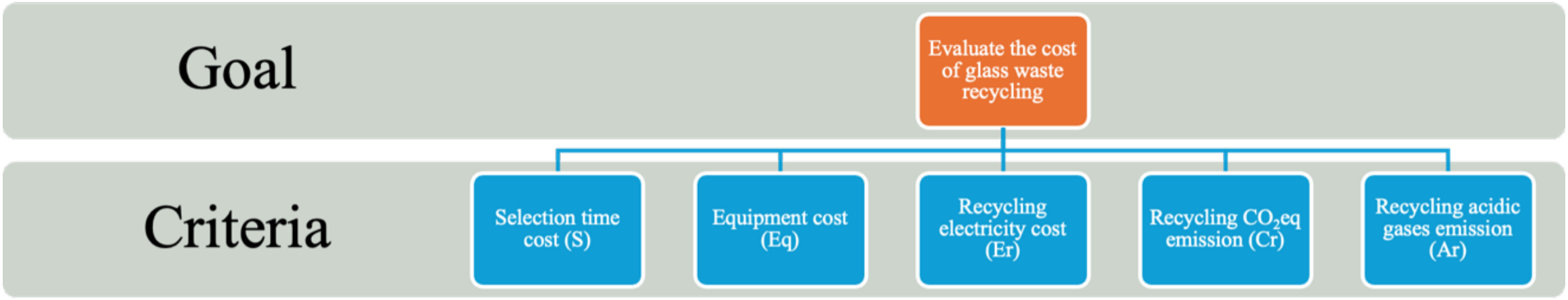

*Fig 3 The Hierarchical Structure of Recycling Cost Model*

### 2. Pairwise Comparison of Criteria

We collected 4 sample questionnaires of the pairwise comparison of criteria from our school teachers (shown in Appendix 1.2), using the AHP 19-point scale. We take the geometric mean of each element in the comparison matrices (here N=4):

$$m_{i,j} = \sqrt[N]{\prod_{k=1}^{N} q_{k_{i,j}}} = \sqrt[4]{\prod_{k=1}^{4} q_{k_{i,j}}} \ (i,j \in [S, Eq, Er, Cr, Ar]) \ (4.15)$$

Then, by computing with a Microsoft Office Excel spreadsheet, we have our final pairwise comparison matrix, rounded to 2 decimal places, with each element $m_{i,j}$ comparing criterion i to criterion j:

$$M = \begin{bmatrix} m_{S,S} & m_{Eq,S} & m_{Er,S} & m_{Cr,S} & m_{Ar,S} \\ m_{S,Eq} & m_{Eq,Eq} & m_{Er,Eq} & m_{Cr,Eq} & m_{Ar,Eq} \\ m_{S,Er} & m_{Eq,Er} & m_{Er,Er} & m_{Cr,Er} & m_{Ar,Er} \\ m_{S,Cr} & m_{Eq,Cr} & m_{Er,Cr} & m_{Cr,Cr} & m_{Ar,Cr} \\ m_{S,Ar} & m_{Eq,Ar} & m_{Er,Ar} & m_{Cr,Ar} & m_{Ar,Ar} \end{bmatrix} = \begin{bmatrix} 1.0000 & 0.7401 & 1.4142 & 2.7108 & 2.9130 \\ 1.3512 & 1.0000 & 2.0000 & 2.2795 & 2.5900 \\ 0.7071 & 0.5000 & 1.0000 & 1.4142 & 1.2779 \\ 0.3689 & 0.4387 & 0.7071 & 1.0000 & 0.3761 \\ 0.3433 & 0.3861 & 0.7825 & 2.6591 & 1.0000 \end{bmatrix} \ (4.16)$$

The matrix theoretically satisfies the diagonally symmetrical property, but not perfectly because of estimation, just as we've discussed before in the previous section:

$$m_{i,j} = \frac{1}{m_{j,i}}, and\ m_{i,i} = 1\ for\ all\ i,j \in [S, Eq, Er, Cr, Ar] \ (4.17)$$

The following linear algebra computations in steps 3 and 4 are performed using the MATLAB program in Appendix 2.2, applying input 2. A detailed explanation of this AHP weight calculation algorithm (we will still use the eigenvector method) is provided in the previous section, so we will not be repeating the details too much in here.

### 3. Consistency Check

By computations with MATLAB, the principal eigenvalue $\lambda_{max}$=5.1593, the consistency index CI=0.0398, the random consistency index RI=1.12 (as the dimension n=5), and the consistency ratio CR=0.0355. Because CR<0.1, we can conclude that this pairwise comparison matrix is considered as consistent.[5]

### 4. Weight calculation

After a series of calculations using MATLAB, we have our normalized principal eigenvector as our weight in equation (4.18) , with distribution of weights Fvisualized in Fig 4.

$$W = [0.2780, 0.3139, 0.1634, 0.0987, 0.1460]^T (4.18)$$

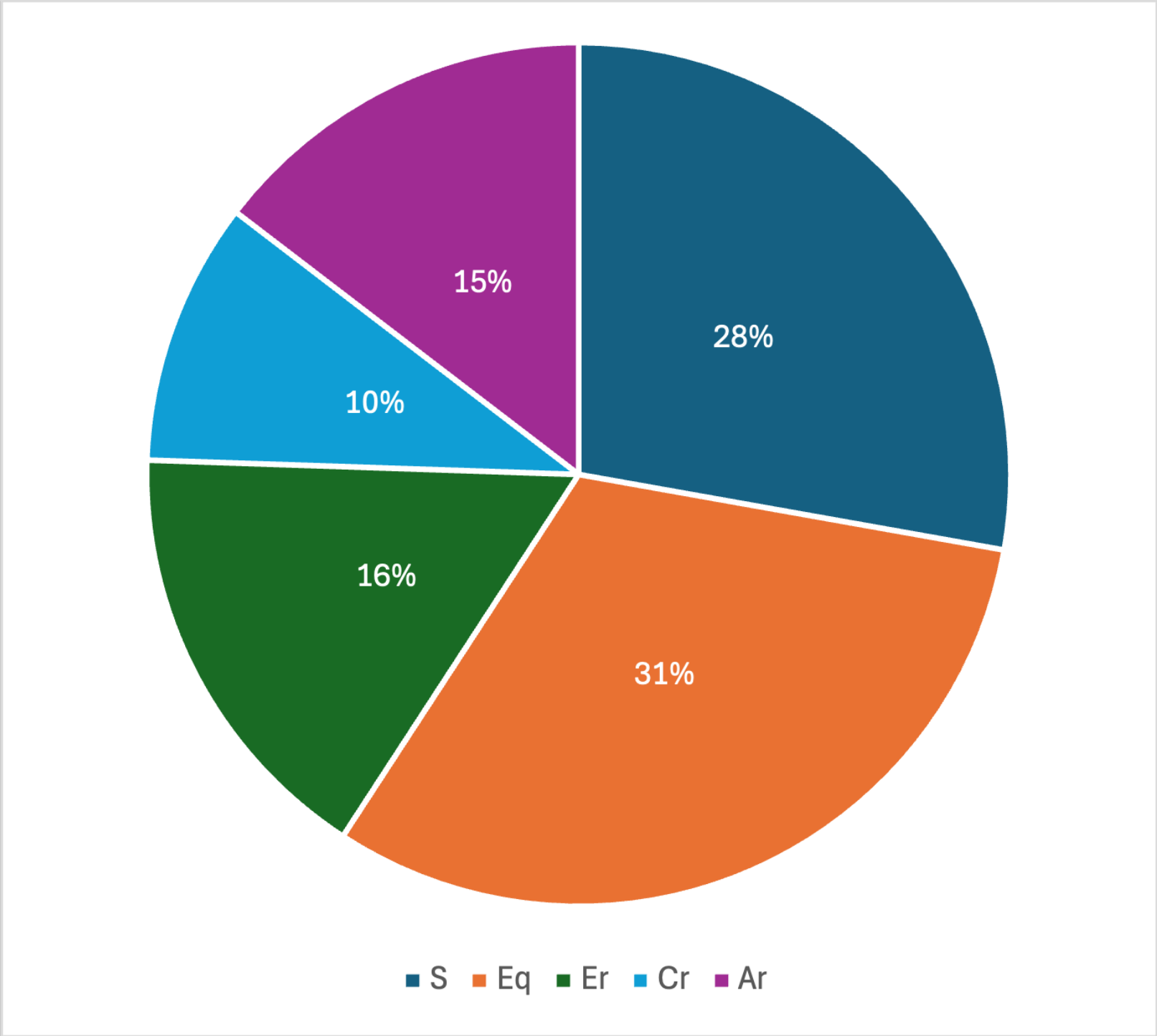

***Fig 4 The Weight Distribution of the Criteria of Glass Recycling Cost Model***

## 4.2.4. Criterion-wise Scoring

Again, we define the row vector of the scores of each criterion in this model:

$$Cri_{glass} = [S_{glass}, Eq_{glass}, Er_{glass}, Cr_{glass}, Ar_{glass}] (4.19)$$

We can obtain values of these data by researching on the Internet, including news reports and official government statistics documents. According to our research: The average volume of a glass bottle is 750cm$^3$, by looking at the standard size a wine container[25], and its reciprocal is 1/750. The price of a big glass recycling treatment machine is about ¥1,500,000 [26]. The energy consumption of recycling glass is approximately 484 kWh/ton[27]. As glass recycling process doesn't have additional air pollution, we can directly calculate by the electricity carbon emission factor in China: 582 g $Co_2$eq / kWh in 2023[28]. So, the carbon emission is $582 \times 484 = 281688g/kWh = 282kg\ /kWh$. There isn't significant emission of acidic gases emission in the recycling process of glass, so this equals 0.

After all, we have our criterion scoring results:

$$Cri_{glass} = \left[\frac{1}{750}, 1500000, 484, 282, 0\right] (4.20)$$

We will perform the normalization of this data together with other types of waste in section 5, after that we can do the calculation of the final cost score by adding weights:

$$CS_{glass} = \sum_{i=1}^{5} Cri_i \cdot W_i \ (4.21)$$

# 5. Generalization of the Model to Multiple Materials

Now, we will generalize this scoring model to various types of waste in Shanghai, adapting the models we developed to calculate their benefits and costs, with an additional consideration of the dynamic relationship between them. We will finally use the scores obtained by the AHP process to reach out an optimized trash recycling collection decision for Shanghai, by recommending 3 types of trash to recycle with priority. The following progress, as an evaluation and optimization model[80], is shown in the MATLAB code of Appendix 2.2.

## 5.1. Adaptation and Interpretation of Benefit Model

The treatment process of different types of trash is quite different, depending on their various property, which is a new consideration that should be taken into account when considering multiple materials. We will continue using the Analytical Hierarchy Process for this model.

### 5.1.1. Re-Building Hierarchical Structure

In the treatment process of some types of wastes, not only greenhouse gases is released, some other pollutants such as acidic gases (causing acidic rains) and toxic gases are also released during the burning treatment. This doesn't exist in the treatment of glass wastes as they are non-burnable materials. It is necessary to add this new criterion At, measured in kg/ton, representing the acidic and toxic gas emission of waste treatment, to the hierarchical structure. Beyond that, we also have to add a new level to our hierarchical structure: the alternatives. The generalized model should now have 6 alternatives: glass (bottles and jars), aluminum (cans and foil), rigid plastic, paper (newspapers, magazines, mixed paper), cardboard, and steel cans. A re-constructed of the hierarchy is shown in Fig 5.

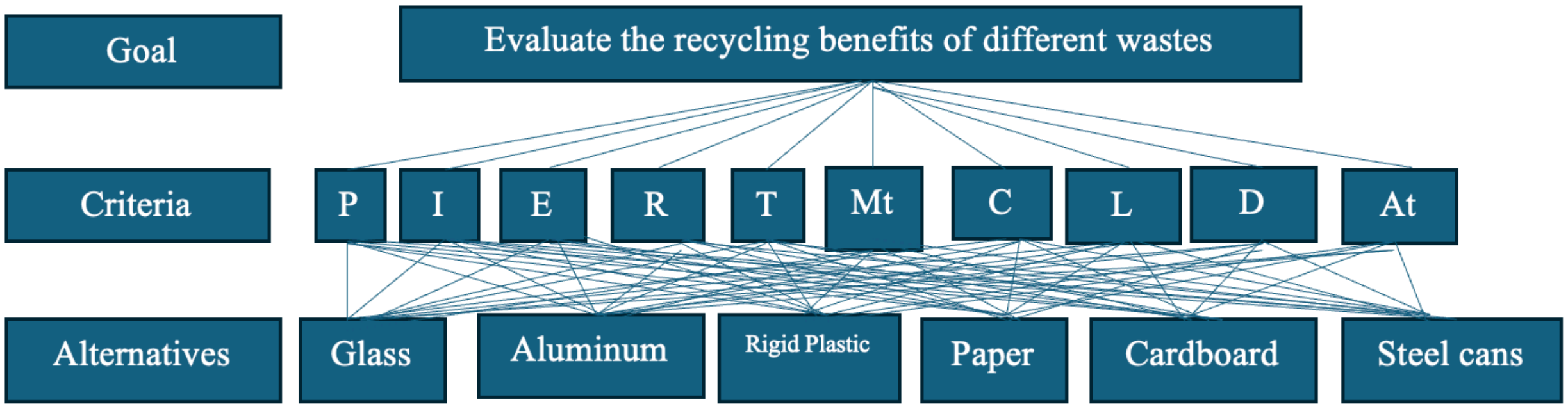

*Fig 5 The Hierarchical Strucutre of Generalized Waste Recycling Benefit Model*

## 5.1.2. Consistency Check & Weight Calculation with AHP process

### Step 1: Build Pairwise Comparison Matrix

We collected the extended pairwise comparison matrices with the criterion At added, by 4 questionnaire samples from our school teachers (in Appendix 1.3.). By taking the geometric mean of each element:

$$m_{i,j} = \sqrt[11]{\prod_{k=1}^{11} q_{k,i,j}} \quad (5.1)$$

We have our average pairwise comparison matrix of the generalized model, rounded to 4 decimal places:

(5.2)

$$M = \begin{bmatrix} m_{P,P} & m_{U,P} & m_{I,P} & m_{E,P} & m_{R,P} & m_{T,P} & m_{Mt,P} & m_{C,P} & m_{L,P} & m_{D,P} & m_{At,P} \\ m_{P,U} & m_{U,U} & m_{I,U} & m_{E,U} & m_{R,U} & m_{T,U} & m_{Mt,U} & m_{C,U} & m_{L,U} & m_{D,U} & m_{At,U} \\ m_{P,I} & m_{U,I} & m_{I,I} & m_{E,I} & m_{R,I} & m_{T,I} & m_{Mt,I} & m_{C,I} & m_{L,I} & m_{D,I} & m_{At,I} \\ m_{P,E} & m_{U,E} & m_{I,E} & m_{E,E} & m_{R,E} & m_{T,E} & m_{Mt,E} & m_{C,E} & m_{L,E} & m_{D,E} & m_{At,E} \\ m_{P,R} & m_{U,R} & m_{I,R} & m_{E,R} & m_{R,R} & m_{T,R} & m_{Mt,R} & m_{C,R} & m_{L,R} & m_{D,R} & m_{At,R} \\ m_{P,T} & m_{U,T} & m_{I,T} & m_{E,T} & m_{R,T} & m_{T,T} & m_{Mt,T} & m_{C,T} & m_{L,T} & m_{D,T} & m_{At,T} \\ m_{P,Mt} & m_{U,Mt} & m_{I,Mt} & m_{E,Mt} & m_{R,Mt} & m_{T,Mt} & m_{Mt,Mt} & m_{C,Mt} & m_{L,Mt} & m_{D,Mt} & m_{At,Mt} \\ m_{P,C} & m_{U,C} & m_{I,C} & m_{E,C} & m_{R,C} & m_{T,C} & m_{Mt,C} & m_{C,C} & m_{L,C} & m_{D,C} & m_{At,C} \\ m_{P,L} & m_{U,L} & m_{I,L} & m_{E,L} & m_{R,L} & m_{T,L} & m_{Mt,L} & m_{C,L} & m_{L,L} & m_{D,L} & m_{At,L} \\ m_{P,D} & m_{U,D} & m_{I,D} & m_{E,D} & m_{R,D} & m_{T,D} & m_{Mt,D} & m_{C,D} & m_{L,D} & m_{D,D} & m_{At,D} \\ m_{P,At} & m_{U,At} & m_{I,At} & m_{E,At} & m_{R,At} & m_{T,At} & m_{Mt,At} & m_{C,At} & m_{L,At} & m_{D,At} & m_{At,At} \end{bmatrix}$$

$$= \begin{bmatrix} 1.0000 & 0.7071 & 2.7832 & 2.1407 & 0.8409 & 3.9843 & 2.2795 & 0.5774 & 1.6226 & 3.1179 & 3.9843 \\ 1.4142 & 1.0000 & 1.5651 & 1.8612 & 1.1067 & 1.5137 & 1.3161 & 0.5373 & 2.7108 & 1.8612 & 3.3098 \\ 0.3593 & 0.6389 & 1.0000 & 0.5466 & 0.5411 & 1.5651 & 2.3784 & 0.5318 & 0.9306 & 1.9343 & 1.6069 \\ 0.4671 & 0.5373 & 1.8294 & 1.0000 & 1.5651 & 1.7321 & 1.4142 & 0.5000 & 1.1892 & 2.4746 & 1.8612 \\ 1.1892 & 0.9036 & 1.8481 & 0.6389 & 1.0000 & 1.5651 & 1.2574 & 0.6606 & 1.0393 & 2.3784 & 2.4495 \\ 0.2510 & 0.6606 & 0.6389 & 0.5774 & 0.6389 & 1.0000 & 1.0000 & 0.4518 & 0.9036 & 1.2574 & 0.4671 \\ 0.4387 & 0.7598 & 0.4204 & 0.7071 & 0.7953 & 1.0000 & 1.0000 & 0.3593 & 0.8409 & 1.7321 & 1.0000 \\ 1.7321 & 1.8612 & 1.8803 & 2.0000 & 1.5137 & 2.2134 & 2.7832 & 1.0000 & 2.7108 & 2.7108 & 2.5149 \\ 0.6148 & 0.3689 & 1.0746 & 0.8409 & 0.9622 & 1.1067 & 1.1892 & 0.3689 & 1.0000 & 2.1407 & 1.7602 \\ 0.3207 & 0.5373 & 0.5170 & 0.4041 & 0.4204 & 0.7953 & 0.5774 & 0.3689 & 0.4671 & 1.0000 & 0.6148 \\ 0.2510 & 0.3021 & 0.6223 & 0.5373 & 0.4082 & 2.1407 & 1.0000 & 0.3976 & 0.5681 & 1.6266 & 1.0000 \end{bmatrix}$$

### Step 2: Consistency test

After a series of computations by MATLAB, the principal eigenvalue $\lambda_{max}$=11.4623, the consistency index CI=0.0462, the random consistency index RI = 1.51 (as dimension n=11), and the consistency ratio CR=0.0306<0.1, so the matrix is considered as consistent.[5]

### Step 3: Weight Calculation

We will still use the eigenvector method to calculate the weight vector. As we've described the algorithm before in the previous section, we don't repeat explaining the details anymore here. After computing with MATLAB, the normalized principal eigenvector derives the weights of each criterion, rounded to 4 decimal places in equation (5.3), with weight distribution visualized in the pie chart in Fig 6.

$$W = [0.1452, 0.1272, 0.0765, 0.0953, 0.1027, 0.0547, 0.0618, 0.1650, 0.0745, 0.0425, 0.0546]^T \quad (5.3)$$

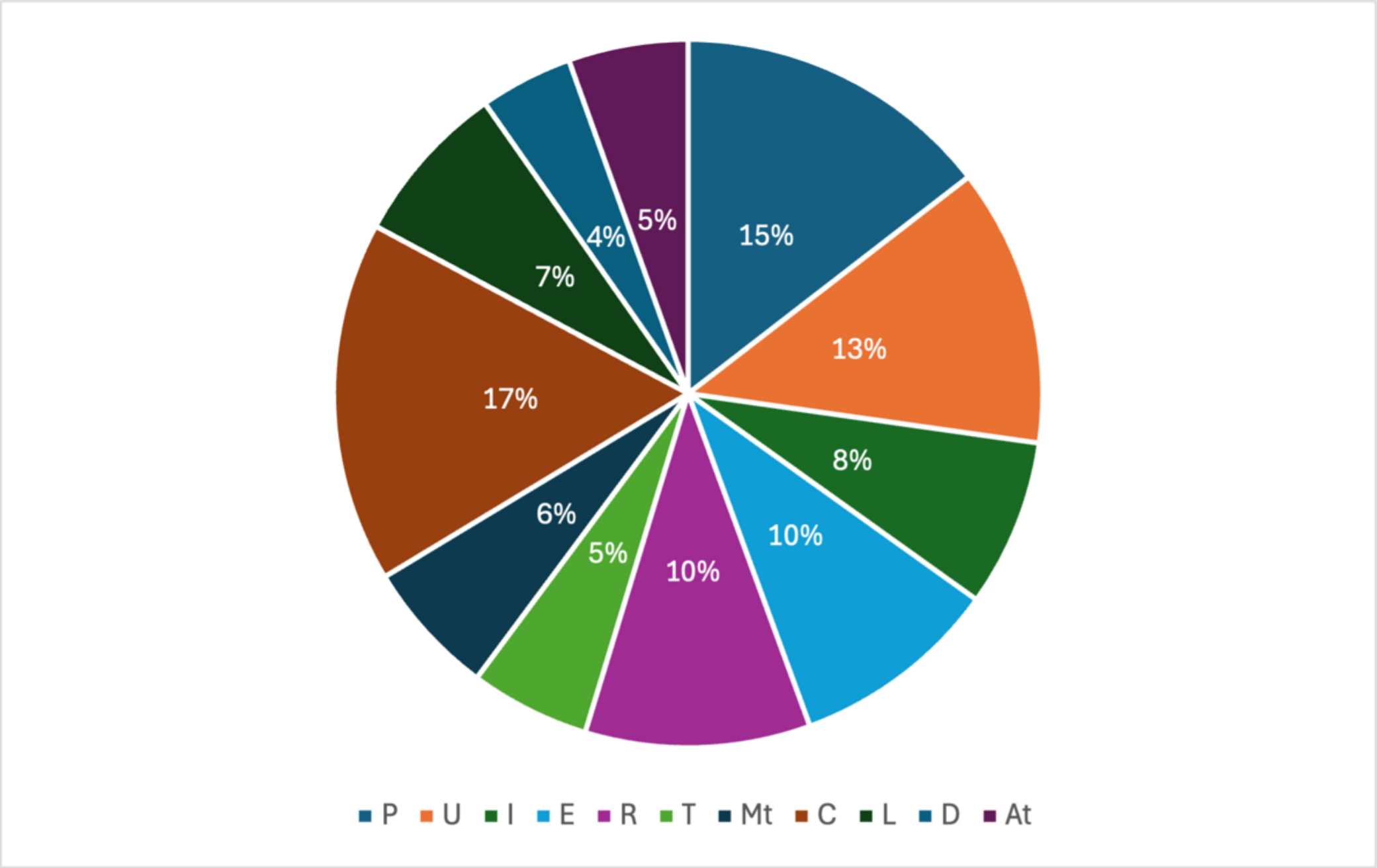

*Fig 6 The Weight Distribution Graph of Generalized Recycling Benefit Model*

## 5.1.3. Criterion-wise Scoring

Define the scores of each criterion j for each type of waste i as a matrix:

$$Cri = [Cri_{glass}, Cri_{alumninum}, Cri_{rigidplastic}, Cri_{paper}, Cri_{cardboard}, Cri_{steelcan}] (5.4)$$

$$Cri_i = [P_i, U_i, I_i, E_i, R_i, T_i, Mt_i, C_i, L_i, D_i, At_i] (5.5)$$

We will collect the data from direct statistics or with an evaluation method. The data of glass was already collected in the previous section with equation (4.13), we only need to assess the additional criterion At added in the section for glass.

### Type 1: Directly Measured Scores

**1) Glass (Bottles and Jars):** Using our common sense, we know that there's 0 acidic or poisonous gas emission during the regular treatment: landfilling of glass wastes.

**2) Aluminum (Cans and foils):** The price of aluminum cans and foil in China is 15500 CN¥/ton in 2024[29]. The usage demand of aluminum cans and foils in China is 4559 kilotons/year[30]. The energy consumption of original production is 17000 kWh/ton[31]=61.2 GJ/ton. The recycling rate of aluminum cans is 73%[32]. The carbon emission of producing aluminum in China 2020, is 14400 kg/t[33]. The density of aluminum is 2710 kg/$m^3$[34], so the space taken up is 369 $cm^3$/ton. The natural decomposition period of aluminum is 200 years[35].

**3) Rigid Plastic:** Rigid plastic mainly include PC, PVC, PP. We will calculate the arithmetic mean of the 3 types for each criterion scoring if a general data for all plastics in not found precisely. The average price of rigid plastic in China is (7500[36]+7000[37] +9000[38])/3=7833 CN¥/ton. In 2021 China's plastic demand is 56350 kilotons[39]. The average energy consumption of rigid plastic is (98+66+88)/3=84 GJ/t[40]. The recycling rate is approximately 80% as the loss rate is 20%[41]. The average carbon emission of producing rigid plastic in China is 3120kg$CO_2$eq/ton[42].The density of plastic is 1 g/$cm^3$, so the space taken up is approximately 1000 $cm^3$/kg[43]. The average natural decomposition period of rigid plastic is 100-500 years[44], taking the average we have 300 years.

**4) Paper (newspapers, magazines, mixed paper):** The price of waste papers is approximately 1000 CN¥/t[45]. The demand of paper in China, 2022 is 280,000 kilotons[46]. The energy consumption of paper production is 11.5 GJ/t[47]. 1 ton of waste paper can be recycled into 0.8 ton of recycled paper[48], so the recycling rate is 80%. The carbon emission of paper production in China is approximately 1345 kg/t[49]. The standard density of paper is 1.201 kg/$m^3$[50], taking up space 832639 $cm^3$/kg. The natural decomposition period of paper is averagely 4 weeks[51], equals approximately 0.083 years.

**5) Cardboard:** The price of cardboard in China, 2024 is 2918 CN¥/t[52]. The demand of carboard is 124020 kilotons in China 2022[53]. The energy consumption of cardboard production is similar to that of paper, which is 11.5 MJ/t. So are the recycling rate, and material primary production carbon emission: 80% and 1345 kg/t. The density of cardboard is 689 kg/$m^3$[54], so the space taken up is approximately 14$cm^3$. The natural decomposition of cardboard typically takes 2 months[55], which equals approximately 0.167 years.

**6) Steel Cans:** The price of tinplate (the most common type of steel to make cans) in China is approximately 3500CN¥/t[56]. The demand of tinplate in China is averagely 8500 kiloton/year[57]. The energy consumption of tinplate production is 49.8 GJ/t[58]. As metal, tinplate can be almost infinitely recycled without loss in quality[59], so the recycle rate is 100%. The carbon emission of primary production of tinplate is 2850 kg $CO_2$eq/ton[60]. The density of tinplate is 7.3 g/$cm^3$[61], so the space taken up is 137 $cm^3$/kg. The natural decomposition period of tinplate is approximately 50 years[62].

## Type 2: Scores with Specific Evaluation Methods

Variables I, T, C and At need special evaluations, as they include the consideration of multiple factors included. Below are the specialized detailed consideration and discussion of the mathematical treatment to these variables.

**1) Limitedness (I):** Although we considered this score in a way of weighted average, in this time, after examining all the wastes, we can see that these are all products made in single material, so here the limitedness is directly the demand of that material divided by the known storage in China (L=D/R). So, the limitedness of aluminum is 4559kilotons/year[30] / 675530 kilotons[63] ≈ 0.0067. The limitedness of petroleum, the main material to manufacture rigid plastics is 764,000 kilotons/year[64] / 3,850,000 kilotons[65] ≈ 0.1984. The limitedness of wood, the material for paper and carboard production, is 15424.59 $m^3$/year[66] / 17,560,000,000 $m^3$[67] ≈0.00000088. The limitedness of steel, the main material of steel cans, is 910,000 kilotons/year[68] / 16,246,000[69] kilotons ≈ 0.056.

**2) Treatment Carbon Emission (T), Treatment Acidic & Toxic gas emission (At):** due to that there 2 regular ways to treat some of the wastes: burning and landfilling, we take the arithmetic mean of the 2 possible values for each criterion. As the wastes here are all non-organic, the carbon and acidic gases emissions are both 0 during landfilling. If there are toxic gases released, we set At to 1000, to show the seriousness and contrast in our evaluation. The consideration for each type of waste is shown below:

**Aluminum:** Burning is not applicable, then T=0, At=0 kg/ton.

**Rigid Plastic:** Burning and landfilling are both applicable, with average carbon emission 1000 kg/t[70], then the average carbon emission T=(1000+0)/2=500. Burning of rigid plastic also release the toxic substance pCCD, so we set T=1000.

**Paper & Cardboard** (share similar property on aspect and burning and landfilling): burning treatment of paper isn't anymore common in China , so we consider only landfilling, with both carbon and toxic gases emission equal to 0.

**Steel cans:** Burning is also not applicable, so T=At=0 kg/ton

## Type 3: Constant Data

The value of C, the cost of landfilling of wastes is the same among all kinds of wastes, as it's not relative to the property of the waste. The average price is just the same as we previously collected for glass: 250 CN¥/t.

# 5.1.4. Normalization and Final Scoring

Now we obtained the criterion-wise scoring matrix in Table 4, by applying the formulae and methodology we derived from AHP process. The calculation was done by Microsoft Excel spreadsheet.

***Table 4 Waste Recycling Benefits Criterion-wise Scoring Matrix***

| | P | U | I | E | R | T | Mt | C | L | D | At |
|---|---|---|---|---|---|---|---|---|---|---|---|
| Glass (bottles, jars) | 1782.3 | 31464 | 0.0089 | 7.2 | 1 | 0 | 1870 | 250 | 370 | 4000 | 0 |
| Aluminum (can, foil) | 15500 | 4559 | 0.0067 | 61.2 | 0.7 | 500 | 14400 | 250 | 369 | 200 | 0 |
| Rigid plastic | 7833 | 56350 | 0.1984 | 84 | 0.8 | 0 | 3120 | 250 | 1000 | 300 | 1000 |
| Paper (newspapers, magazines, mixed paper ) | 1000 | 280000 | 0.00000088 | 11.5 | 0.8 | 0 | 1345 | 250 | 832639 | 0.083 | 0 |
| Cardboard | 2918 | 124020 | 0.00000088 | 11.5 | 0.8 | 0 | 1345 | 250 | 1451 | 0.167 | 0 |
| Steel cans | 3500 | 8500 | 0.056 | 49.8 | 1 | 0 | 2850 | 250 | 137 | 50 | 0 |

Before we evaluate the score, it's essential to normalize the data for each criterion. Here we will use the min-max normalization algorithm:

$$x' = \frac{x - \min(x)}{\max(x) - \min(x)} \quad (5.6)$$

Applying in case of this matrix:

$$Cri'_{i,j} = \frac{Cri_{i,j} - \min(Cri_i)}{\max(Cri_i) - \min(Cri_i)} \quad (\max(Cri_i) - \min(Cri_i) \neq 0)$$

$$if\ \max(Cri_i) - \min(Cri_i) = 0, then\ Cri_{i,j} = 1\ for\ all\ j \quad (5.7)$$

Computing with MATLAB, we have our normalized scoring matrix, rounded to 4 decimal places, as shown in Table 5.

***Table 5 Normalized Waste Recycling Benefits Scoring Matrix***

| | P | U | I | E | R | T | Mt | C | L | D | At |
|---|---|---|---|---|---|---|---|---|---|---|---|
| Glass (bottles, jars) | 0.0540 | 0.0977 | 0.0449 | 0.0000 | 1.0000 | 0.0000 | 0.0402 | 1.0000 | 0.0003 | 1.0000 | 0.0000 |
| Aluminum (can, foil) | 1.0000 | 0.0000 | 0.0338 | 0.7031 | 0.0000 | 1.0000 | 1.0000 | 1.0000 | 0.0003 | 0.0500 | 0.0000 |
| Rigid plastic | 0.4712 | 0.1880 | 1.0000 | 1.0000 | 0.2593 | 0.0000 | 0.1360 | 1.0000 | 0.0010 | 0.0750 | 1.0000 |
| Paper (newspapers, magazines, mixed paper ) | 0.0000 | 1.0000 | 0.0000 | 0.0560 | 0.2593 | 0.0000 | 0.0000 | 1.0000 | 1.0000 | 0.0000 | 0.0000 |
| Cardboard | 0.1323 | 0.4337 | 0.0000 | 0.0560 | 0.2593 | 0.0000 | 0.0000 | 1.0000 | 0.0016 | 0.0000 | 0.0000 |
| Steel cans | 0.1724 | 0.0143 | 0.2823 | 0.5547 | 1.0000 | 0.0000 | 0.1153 | 1.0000 | 0.0000 | 0.0125 | 0.0000 |

We apply the weight adding formula:

$$BS_j = \sum_{i=1}^{10} NormalizedCri_{j,i} \cdot W_i \quad for\ all\ j \quad (5.8)$$

With the weight vector:

$$W = [0.1452, 0.1272, 0.0765, 0.0953, 0.1027, 0.0547, 0.0618, 0.1650, 0.0745, 0.0425, 0.0546]^T \quad (5.9)$$

By MATLAB computations, we have our benefit score vector for each type of trash, in Table 6. By conclusion, the quantified benefit of glass recycling is 0.3364.

***Table 6 The Benefit Scores of Recycling of Different Types of Trash***

| Type of Waste | BS |
|---|---|
| Glass (bottles, jars) | 0.3364 |
| Aluminum (can, foil) | 0.4984 |
| Rigid plastic | 0.5220 |
| Paper (newspapers, magazines, mixed paper ) | 0.3986 |
| Cardboard | 0.2714 |
| Steel cans | 0.3766 |

## 5.2. Adaptation and Interpretation of Cost Model

There're no essential extra assessment criteria to add in the cost model, so we can directly proceed to data collection, with all data taken directly without complex pre-procession from Internet research statistics[71][72][73][74][75]. As this process is quite simple, so we don't explain it repetitively in here. For carbon emission, we multiply the electricity consumption by the carbon emission factor 582 g $Co_2eq$ / kWh in China. 0 acidic gases are released during the recycle process. The final criterion-wise scoring is shown below, in Table 7. After min-max normalization, we have the normalized Cri matrix in Table 8.

***Table 7 The Criterion-wise Scoring Matrix of Waste Recycling Costs Model***

| | S | Eq | Er | Cr | Ar |
|---|---|---|---|---|---|
| Glass (bottles, jars) | 0.0013 | 1500000.0000 | 831.6151 | 484.0000 | 0.0000 |
| Aluminum (can, foil) | 0.0019 | 68000.0000 | 13500.0000 | 7857.0000 | 0.0000 |
| Rigid plastic | 0.2000 | 4853.0000 | 300.0000 | 174.6000 | 0.0000 |
| Paper (newspapers, magazines, mixed paper ) | 0.0300 | 42000.0000 | 120.2749 | 70.0000 | 0.0000 |
| Cardboard | 0.0007 | 42000.0000 | 120.2749 | 70.0000 | 0.0000 |
| Steel cans | 0.0019 | 68000.0000 | 13500.0000 | 7857.0000 | 0.0000 |

***Table 8 Normalized Waste Recycling Cost Criterion-wise Scoring Matrix***

| | S | Eq | Er | Cr | Ar |
|---|---|---|---|---|---|
| Glass (bottles, jars) | 0.0030 | 1.0000 | 0.0532 | 0.0532 | 1.0000 |
| Aluminum (can, foil) | 0.0060 | 0.0422 | 1.0000 | 1.0000 | 1.0000 |
| Rigid plastic | 1.0000 | 0.0000 | 0.0134 | 0.0134 | 1.0000 |
| Paper (newspapers, magazines, mixed paper ) | 0.1470 | 0.0248 | 0.0000 | 0.0000 | 1.0000 |
| Cardboard | 0.0000 | 0.0248 | 0.0000 | 0.0000 | 1.0000 |
| Steel cans | 0.0060 | 0.0422 | 1.0000 | 1.0000 | 1.0000 |

Then, we apply the weight-adding formula:

$$Cri = [Cri_{glass}, Cri_{alumninum}, Cri_{rigidplastic}, Cri_{paper}, Cri_{cardboard}, Cri_{steelcan}] (5.9)$$

$$Cri_i = [P_i, U_i, I_i, E_i, R_i, T_i, Mt_i, C_i, L_i, D_i, At_i] (5.10)$$

With the weight vector we calculated in the previous section:

$$W = [0.2780, 0.3139, 0.1634, 0.0987, 0.1460]^T \ (5.11)$$

This yields the final cost score vector, as shown in Table 9. With completion to the glass recycling model in the previous section, we have a quantified cost of glass recycling as 0.4746.

*Table 9 The Cost Scores of Different Types of Wastes*

| Trash Type | CS |
|---|---|
| Glass (bottles, jars) | 0.4746 |
| Aluminum (can, foil) | 0.4230 |
| Rigid plastic | 0.4275 |
| Paper (newspapers, magazines, mixed paper ) | 0.1946 |
| Cardboard | 0.1538 |
| Steel cans | 0.4230 |

## 5.3. Interdependence Consideration: Transportation, Contamination and Co-Pollution Costs

To improve the robustness and reasonability of our model, we have to discuss the following further considerations. As the 3 types of trash we're going to choose is going to be in transported the same truck, so they'll have a dynamic effect on each other as our decision varies. The contamination cost is actually already included in our AHP model, with the criterion selection time cost (S). The co-pollution cost consideration is actually non-essential for this model, as the recycling of each trash isn't affected by the other's pollution – and as they are all non-organic materials, and the organic kitchen wastes are always previously separated from the dry wastes in Shanghai, so there isn't such a serious pollution to consider with. The transportation cost of different materials is approximately the same, the only difference in the space they take up in the limited area of truck. However, this is just perfectly reflected by the variable L, the space the trash takes up per unit mass, which is already in the consideration of our assessment criteria, so there's no further need to extend our model here.

## 5.4. The Waste Recycling Evaluation Model

Afterall, with all the considerations contained in our model, we can now calculate our final recycling evaluation score as the difference of the benefit and cost:

$$FS_i = BS_i - CS_i \quad (5.12)$$

We finally have our final scores (Table 10), as an index to compare the benefit and cost of the recycling process of each type of waste in Shanghai. So, our final optimized decision is to prioritize the recycling of the wastes with a higher FS, as shown in Table 10.

*Table 10 The Final Evaluation Scores*

| Trash Type | FS |
|---|---|
| Glass (bottles, jars) | -0.13827331 |
| Aluminum (can, foil) | 0.075331369 |
| Rigid plastic | 0.094564818 |
| Paper (newpapers, magazines, mixed paper ) | 0.203986922 |
| Cardboard | 0.117649746 |
| Steel cans | -0.04641927 |

# 6. Evaluation of Model

## 6.1. Strengths

There are several strengths of our model. As we can see, we used the professional mathematical evaluation process: the Analytical Hierarchy Process (AHP), which is good and appropriate to reach out a crucial result in this situation of usage. During the process, we have used crucial methods to do the computations, including the min-max normalization process, the weighted sum, the geometric mean, the eigenvector and the eigenvalue. These methods help us to build a completely logical and comprehensive model, with classical algorithms developed by Sir Thomas Saaty, considered as the most appropriate model to determine the weights of an evaluation through subjective choices, with both significance in study of psychology and mathematics.

Before this process, we also had several assumptions and pre-definitions to ensure the accuracy of our model. A variable table is made to organize the notations used throughout the model in a clear way. We used the computer algebra system MATLAB to perform all the computations involved in the model, this highly improves the accuracy of the values, as the computation processes such as finding the eigenvector involves very complex linear algebra and differentiation, and using a software is much more reliable than hands-on calculation. Our data collection process is also very crucial, with each data point accompanied with the corresponding reference, and logical explanations, and possibly simple pre-procession calculations.

## 6.2. Weaknesses

However, we have one points to impArove in our model. The dynamic relationship between the amount of each trash and the optimization of the others isn't shown as perfectly optimized in this model. Maybe we can try to structure a set of relative functions of each trash on the other, and use the techniques of finding partial derivatives, solve the simultaneous partial differential equations, and then find a totally dynamically optimized set of solution

# 7. Conclusion

In this research, we applied mathematical methodologies based on the classical Analytical Hierarchy Process method and provided a rigorous analysis on the benefits and costs of the recycling process of different types of household wastes in Shanghai. From the model, we derived a list of indices determining the priority of recycling on each type of waste, according to the benefits gained and the costs applied. The optimization method will provide a logical decision-making on the optimization of a city's waste recycling system, accelerating the process of environment protection. Our whole work showcases an integration between mathematical modeling and sustainability developments, showing the possibility of further cooperations of these 2 distinct fields of study.

# Acknowledgements

As according to the policy of the International Mathematical Modeling Challenge (Greater China) Committee, herewith I specify that this is a paper submitted for the IMMC competition, with a research objective of optimizing some city's recycling program which is from the problem set out by the Problem Setting Committee of IMMC (Greater China). The research methodology and all other academic works in this essay are all originally created by the authors.

# Appendices

## Appendix I. Questionnaire Sample Datasets of Pairwise Comparison Matrices

### Appendix 1.1. Questionnaire Samples Pairwise Comparison Matrices of Glass Waste Recycling Benefit Model

$$Q = [q_1, q_2, q_3, q_4]$$

$$q_1 = \begin{bmatrix} 1 & 1 & 5 & 7 & 1 & 4 & 3 & 1 & 2 & 7 \\ 1 & 1 & 1 & 3 & 3 & \frac{1}{4} & \frac{1}{4} & \frac{1}{6} & 2 & 3 \\ \frac{1}{5} & 1 & 1 & \frac{1}{7} & \frac{1}{7} & 1 & 2 & \frac{1}{5} & \frac{1}{4} & 4 \\ \frac{1}{7} & \frac{1}{3} & 7 & 1 & 6 & 1 & 2 & \frac{1}{4} & \frac{1}{5} & 5 \\ 1 & \frac{1}{3} & 7 & \frac{1}{6} & 1 & \frac{1}{3} & \frac{1}{2} & \frac{1}{7} & \frac{1}{6} & 2 \\ \frac{1}{4} & 4 & 1 & 1 & 3 & 1 & 1 & \frac{1}{3} & \frac{1}{3} & 5 \\ \frac{1}{3} & 4 & \frac{1}{2} & \frac{1}{2} & 2 & 1 & 1 & \frac{1}{5} & \frac{1}{4} & 3 \\ 1 & 6 & 5 & 4 & 7 & 3 & 5 & 1 & 3 & 6 \\ \frac{1}{2} & \frac{1}{2} & 4 & 5 & 6 & 3 & 4 & \frac{1}{3} & 1 & 7 \\ \frac{1}{7} & \frac{1}{3} & \frac{1}{4} & \frac{1}{5} & \frac{1}{2} & \frac{1}{5} & \frac{1}{3} & \frac{1}{6} & \frac{1}{7} & 1 \end{bmatrix} \quad q_2 = \begin{bmatrix} 1 & \frac{1}{4} & 2 & 1 & \frac{1}{2} & 3 & 1 & 1 & 1 & \frac{1}{2} \\ 4 & 1 & 1 & 1 & \frac{1}{2} & 3 & 3 & 1 & 1 & 1 \\ \frac{1}{2} & 1 & 1 & \frac{1}{4} & 1 & 2 & 2 & 1 & 1 & \frac{1}{2} \\ 1 & 1 & 4 & 1 & 1 & 1 & 1 & 1 & 1 & \frac{1}{2} \\ 2 & 2 & 1 & 1 & 1 & 3 & 1 & \frac{1}{3} & 2 & 1 \\ \frac{1}{3} & \frac{1}{3} & \frac{1}{2} & 1 & \frac{1}{3} & 1 & 1 & 1 & \frac{1}{2} & \frac{1}{3} \\ 1 & \frac{1}{3} & \frac{1}{2} & 1 & 1 & 1 & 1 & \frac{1}{4} & 1 & \frac{1}{2} \\ 1 & 1 & 1 & 1 & 3 & 1 & 4 & 1 & 1 & 1 \\ 1 & 1 & 1 & 1 & \frac{1}{2} & 2 & 1 & 1 & 1 & 1 \\ 2 & 1 & 2 & 2 & 1 & 3 & 2 & 1 & 1 & 1 \end{bmatrix}$$

$$q_3 = \begin{bmatrix} 1 & \frac{1}{2} & 2 & 3 & 3 & 3 & 3 & \frac{1}{3} & \frac{1}{2} & 3 \\ 2 & 1 & 2 & 4 & 2 & 1 & 1 & 1 & 3 & \frac{1}{2} \\ \frac{1}{2} & \frac{1}{2} & 1 & 5 & 3 & 1 & 4 & 2 & \frac{1}{2} & 1 \\ \frac{1}{3} & \frac{1}{4} & \frac{1}{5} & 1 & 2 & 3 & 2 & \frac{1}{2} & 2 & 3 \\ \frac{1}{3} & \frac{1}{2} & \frac{1}{3} & \frac{1}{2} & 1 & 1 & 1 & 4 & \frac{1}{2} & 2 \\ \frac{1}{3} & 1 & 1 & \frac{1}{3} & 1 & 1 & 2 & \frac{1}{2} & 2 & \frac{1}{2} \\ \frac{1}{3} & 1 & \frac{1}{4} & \frac{1}{2} & 1 & \frac{1}{2} & 1 & 2 & 1 & 2 \\ 3 & 1 & \frac{1}{2} & 2 & \frac{1}{4} & 2 & \frac{1}{2} & 1 & 2 & 1 \\ 2 & \frac{1}{3} & 2 & \frac{1}{2} & 2 & \frac{1}{2} & 1 & \frac{1}{2} & 1 & 1 \\ \frac{1}{3} & 2 & 1 & \frac{1}{3} & \frac{1}{2} & 2 & \frac{1}{2} & 1 & 1 & 1 \end{bmatrix} q_4 = \begin{bmatrix} 1 & 2 & 3 & 1 & \frac{1}{3} & 7 & 3 & \frac{1}{3} & 7 & 9 \\ \frac{1}{2} & 1 & 3 & 1 & \frac{1}{2} & 7 & 4 & \frac{1}{2} & 9 & 8 \\ \frac{1}{3} & \frac{1}{3} & 1 & \frac{1}{2} & \frac{1}{5} & 3 & 2 & \frac{1}{5} & 6 & 7 \\ 1 & 1 & 2 & 1 & \frac{1}{2} & 3 & 1 & \frac{1}{2} & 5 & 5 \\ 3 & 2 & 5 & 2 & 1 & 6 & 5 & 1 & 7 & 8 \\ \frac{1}{7} & \frac{1}{7} & \frac{1}{3} & \frac{1}{3} & \frac{1}{6} & 1 & \frac{1}{2} & \frac{1}{4} & 2 & 3 \\ \frac{1}{3} & \frac{1}{4} & \frac{1}{2} & 1 & \frac{1}{5} & 2 & 1 & \frac{1}{6} & 2 & 3 \\ 3 & 2 & 5 & 2 & 1 & 4 & 6 & 1 & 9 & 9 \\ \frac{1}{7} & \frac{1}{9} & \frac{1}{6} & \frac{1}{5} & \frac{1}{7} & \frac{1}{2} & \frac{1}{2} & \frac{1}{9} & 1 & 3 \\ \frac{1}{9} & \frac{1}{8} & \frac{1}{7} & \frac{1}{5} & \frac{1}{8} & \frac{1}{3} & \frac{1}{3} & \frac{1}{9} & \frac{1}{3} & 1 \end{bmatrix}$$

# Appendix 1.2. Pairwise Comparison Matrices of Glass Waste Recycling Cost Model

$$Q = [q_1, q_2, q_3, q_4]$$

$$q_1 = \begin{bmatrix} 1 & \frac{1}{5} & \frac{1}{3} & 2 & 3 \\ 5 & 1 & 4 & 5 & 6 \\ 3 & \frac{1}{4} & 1 & \frac{1}{3} & 2 \\ \frac{1}{2} & \frac{1}{5} & 3 & 1 & 5 \\ \frac{1}{3} & \frac{1}{6} & \frac{1}{2} & \frac{1}{5} & 1 \end{bmatrix} q_2 = \begin{bmatrix} 1 & 1 & 2 & 2 & 1 \\ 1 & 1 & 1 & 2 & 1 \\ \frac{1}{2} & 1 & 1 & 2 & 2 \\ \frac{1}{2} & \frac{1}{2} & \frac{1}{2} & 1 & 1 \\ 1 & 1 & \frac{1}{2} & 1 & 1 \end{bmatrix}$$

$$q_3 = \begin{bmatrix} 1 & 3 & 2 & 2 & 2 \\ \frac{1}{3} & 1 & 2 & \frac{1}{2} & \frac{1}{2} \\ \frac{1}{2} & \frac{1}{2} & 1 & 2 & \frac{1}{2} \\ \frac{1}{2} & 2 & \frac{1}{2} & 1 & 2 \\ \frac{1}{2} & 2 & 2 & \frac{1}{2} & 1 \end{bmatrix} q_4 = \begin{bmatrix} 1 & \frac{1}{2} & 3 & 9 & 9 \\ 2 & 1 & 2 & 9 & 9 \\ \frac{1}{3} & \frac{1}{2} & 1 & 2 & 2 \\ \frac{1}{9} & \frac{1}{9} & \frac{1}{2} & 1 & \frac{1}{5} \\ \frac{1}{9} & \frac{1}{9} & \frac{1}{2} & 5 & 1 \end{bmatrix}$$

# Appendix 1.3. Pairwise Comparison Matrices of Generalized Recycling Benefit Model

$$Q = [q_1, q_2, q_3, q_4]$$

$$q_1 = \begin{bmatrix} 1 & 1 & 5 & 7 & 1 & 4 & 3 & 1 & 2 & 7 & 9 \\ 1 & 1 & 1 & 3 & 3 & \frac{1}{4} & \frac{1}{4} & \frac{1}{6} & 2 & 3 & 5 \\ \frac{1}{5} & 1 & 1 & \frac{1}{7} & \frac{1}{7} & 1 & 2 & \frac{1}{5} & \frac{1}{4} & 4 & \frac{1}{3} \\ \frac{1}{7} & \frac{1}{3} & 7 & 1 & 6 & 1 & 2 & \frac{1}{4} & \frac{1}{5} & 5 & 3 \\ 1 & \frac{1}{3} & 7 & \frac{1}{6} & 1 & \frac{1}{3} & \frac{1}{2} & \frac{1}{7} & \frac{1}{6} & 2 & 2 \\ \frac{1}{4} & 4 & 1 & 1 & 3 & 1 & 1 & \frac{1}{3} & \frac{1}{3} & 5 & \frac{1}{7} \\ \frac{1}{3} & 4 & \frac{1}{2} & \frac{1}{2} & 2 & 1 & 1 & \frac{1}{5} & \frac{1}{4} & 3 & \frac{1}{2} \\ 1 & 6 & 5 & 4 & 7 & 3 & 5 & 1 & 3 & 6 & 5 \\ \frac{1}{2} & \frac{1}{2} & 4 & 5 & 6 & 3 & 4 & \frac{1}{3} & 1 & 7 & 6 \\ \frac{1}{7} & \frac{1}{3} & \frac{1}{4} & \frac{1}{5} & \frac{1}{2} & \frac{1}{5} & \frac{1}{3} & \frac{1}{6} & \frac{1}{7} & 1 & \frac{1}{4} \\ \frac{1}{9} & \frac{1}{5} & 3 & \frac{1}{3} & \frac{1}{2} & 7 & 2 & \frac{1}{5} & \frac{1}{6} & 4 & 1 \end{bmatrix} q_2 = \begin{bmatrix} 1 & \frac{1}{4} & 2 & 1 & \frac{1}{2} & 3 & 1 & 1 & 1 & \frac{1}{2} & 2 \\ 4 & 1 & 1 & 1 & \frac{1}{2} & 3 & 3 & 1 & 1 & 1 & 3 \\ \frac{1}{2} & 1 & 1 & \frac{1}{4} & 1 & 2 & 2 & 1 & 1 & \frac{1}{2} & 2 \\ 1 & 1 & 4 & 1 & 1 & 1 & 1 & 1 & 1 & \frac{1}{2} & 2 \\ 2 & 2 & 1 & 1 & 1 & 3 & 1 & \frac{1}{3} & 2 & 1 & 1 \\ \frac{1}{3} & \frac{1}{3} & \frac{1}{2} & 1 & \frac{1}{3} & 1 & 1 & 1 & \frac{1}{2} & \frac{1}{3} & 1 \\ 1 & \frac{1}{3} & \frac{1}{2} & 1 & 1 & 1 & 1 & \frac{1}{4} & 1 & \frac{1}{2} & 1 \\ 1 & 1 & 1 & 1 & 3 & 1 & 4 & 1 & 1 & 1 & 1 \\ 1 & 1 & 1 & 1 & \frac{1}{2} & 2 & 1 & 1 & 1 & 1 & 2 \\ 2 & 1 & 2 & 2 & 1 & 3 & 2 & 1 & 1 & 1 & 2 \\ \frac{1}{2} & \frac{1}{3} & \frac{1}{2} & \frac{1}{2} & 1 & 1 & 1 & 1 & \frac{1}{2} & \frac{1}{2} & 1 \end{bmatrix}$$

$$q_3 = \begin{bmatrix} 1 & \frac{1}{2} & 2 & 3 & 3 & 3 & 3 & \frac{1}{3} & \frac{1}{2} & 3 & 2 \\ 2 & 1 & 2 & 4 & 2 & 1 & 1 & 1 & 3 & \frac{1}{2} & 1 \\ \frac{1}{2} & \frac{1}{2} & 1 & 5 & 3 & 1 & 4 & 2 & \frac{1}{2} & 1 & 2 \\ \frac{1}{3} & \frac{1}{4} & \frac{1}{5} & 1 & 2 & 3 & 2 & \frac{1}{2} & 2 & 3 & 1 \\ \frac{1}{3} & \frac{1}{2} & \frac{1}{3} & \frac{1}{2} & 1 & 1 & 1 & 4 & \frac{1}{2} & 2 & 3 \\ \frac{1}{3} & 1 & 1 & \frac{1}{3} & 1 & 1 & 2 & \frac{1}{2} & 2 & \frac{1}{2} & 1 \\ \frac{1}{3} & 1 & \frac{1}{4} & \frac{1}{2} & 1 & \frac{1}{2} & 1 & 2 & 1 & 2 & 1 \\ 3 & 1 & \frac{1}{2} & 2 & \frac{1}{4} & 2 & \frac{1}{2} & 1 & 2 & 1 & 1 \\ 2 & \frac{1}{3} & 2 & \frac{1}{2} & 2 & \frac{1}{2} & 1 & \frac{1}{2} & 1 & 1 & 4 \\ \frac{1}{3} & 2 & 1 & \frac{1}{3} & \frac{1}{2} & 2 & \frac{1}{2} & 1 & 1 & 1 & 2 \\ \frac{1}{2} & 1 & \frac{1}{2} & 1 & \frac{1}{3} & 1 & 1 & 1 & \frac{1}{4} & \frac{1}{2} & 1 \end{bmatrix} q_4 = \begin{bmatrix} 1 & 2 & 3 & 1 & \frac{1}{3} & 7 & 3 & \frac{1}{3} & 7 & 9 & 7 \\ \frac{1}{2} & 1 & 3 & 1 & \frac{1}{2} & 7 & 4 & \frac{1}{2} & 9 & 8 & 8 \\ \frac{1}{3} & \frac{1}{3} & 1 & \frac{1}{2} & \frac{1}{5} & 3 & 2 & \frac{1}{5} & 6 & 7 & 5 \\ 1 & 1 & 2 & 1 & \frac{1}{2} & 3 & 1 & \frac{1}{2} & 5 & 5 & 2 \\ 3 & 2 & 5 & 2 & 1 & 6 & 5 & 1 & 7 & 8 & 6 \\ \frac{1}{7} & \frac{1}{7} & \frac{1}{3} & \frac{1}{3} & \frac{1}{6} & 1 & \frac{1}{2} & \frac{1}{4} & 2 & 3 & \frac{1}{3} \\ \frac{1}{3} & \frac{1}{4} & \frac{1}{2} & 1 & \frac{1}{5} & 2 & 1 & \frac{1}{6} & 2 & 3 & 2 \\ 3 & 2 & 5 & 2 & 1 & 4 & 6 & 1 & 9 & 9 & 8 \\ \frac{1}{7} & \frac{1}{9} & \frac{1}{6} & \frac{1}{5} & \frac{1}{7} & \frac{1}{2} & \frac{1}{2} & \frac{1}{9} & 1 & 3 & \frac{1}{5} \\ \frac{1}{9} & \frac{1}{8} & \frac{1}{7} & \frac{1}{5} & \frac{1}{8} & \frac{1}{3} & \frac{1}{3} & \frac{1}{9} & \frac{1}{3} & 1 & \frac{1}{7} \\ \frac{1}{7} & \frac{1}{8} & \frac{1}{5} & \frac{1}{2} & \frac{1}{6} & 3 & \frac{1}{2} & \frac{1}{8} & 5 & 7 & 1 \end{bmatrix}$$

# Appendix II. MATLAB Codes

Below are the MATLAB codes used for computations throughout the essay.

# Appendix 2.1. Glass Recycling Benefit (Input 1) / Cost (Input 2) Model's AHP Weight Calculation Process

```
% before running the program, please remove one of the inputs and keep the one you need corresponding to the model needed

M=[1.0000   0.7071  2.7832  2.1407  0.8409  3.9843  2.2795  0.5774  1.6266  3.1179
1.4142  1.0000  1.5651  1.8612  1.1067  1.5137  1.3161  0.5373  2.7108  1.8612
0.3593  0.6389  1.0000  0.5466  0.5411  1.5651  2.3784  0.5318  0.9306  1.9343
0.4671  0.5373  1.8294  1.0000  1.5651  1.7321  1.4142  0.5000  1.1892  2.4746
1.1892  0.9036  1.8481  0.6389  1.0000  1.5651  1.2574  0.6606  1.0393  2.3784
0.2510  0.6606  0.6389  0.5774  0.6389  1.0000  1.0000  0.4518  0.9036  1.2574
0.4387  0.7598  0.4204  0.7071  0.7953  1.0000  1.0000  0.3593  0.8409  1.7321
1.7321  1.8612  1.8803  2.0000  1.5137  2.2134  2.7832  1.0000  2.7108  2.7108
0.6148  0.3689  1.0746  0.8409  0.9622  1.1067  1.1892  0.3689  1.0000  2.1407
0.3207  0.5373  0.5170  0.4041  0.4204  0.7953  0.5774  0.3689  0.4671  1.0000]%input 1, glass benefit model criteria pairwise comparison matrix
M=[1.0000   0.7401  1.4142  2.7108  2.9130
1.3512  1.0000  2.0000  2.2795  2.5900
0.7071  0.5000  1.0000  1.4142  1.2779
0.3689  0.4387  0.7071  1.0000  0.3761
0.3433  0.3861  0.7825  2.6591  1.0000] %input 2, glass cost model criteria pairwise comparison matrix
% consistency test
[n,n]=size(M) % get the dimension of M
[V,lambda]=eig(M) % returns diagonal matrix lambda of eigenvalues and matrix V whose columns are the corresponding right eigenvectors
lambda_max=max(max(lambda)) % find the maximum eigenvalue lambda_max
CI=(lambda_max-n)/(n-1) % compute the consistency index (CI) of M
RI=[0.00,0.00,0.58,0.90,1.12,1.24,1.32,1.41,1.45,1.49,1.51,1.48,1.56,1.57]
CR=CI/RI(n)
% calculating weights
[~,c]=find(lambda_max==lambda,1) % find the index of the principal vector (as it's stored in a diagonal matrix, we take the column index)
%The principal eigenvector is then V(:,c)
W=V(:,c)./sum(V(:,c)) % normalization of principal eigenvector

%output 1:W = [0.1491;0.1300;0.0813;0.1017;0.1066;0.0617;0.0668;0.1788;0.0776;0.0464]
%output 2:W =[0.2780; 0.3139; 0.1633; 0.0987; 0.1460]
```

# Appendix 2.2. The Generalized Evaluation and Optimization Model of Trash Recycling in Shanghai

```
%%Part 1: benefit model
M=[1.0000   0.7071  2.7832  2.1407  0.8409  3.9843  2.2795  0.5774  1.6266  3.1179  3.9843
1.4142  1.0000  1.5651  1.8612  1.1067  1.5137  1.3161  0.5373  2.7108  1.8612  3.3098
0.3593  0.6389  1.0000  0.5466  0.5411  1.5651  2.3784  0.5318  0.9306  1.9343  1.6069
0.4671  0.5373  1.8294  1.0000  1.5651  1.7321  1.4142  0.5000  1.1892  2.4746  1.8612
1.1892  0.9036  1.8481  0.6389  1.0000  1.5651  1.2574  0.6606  1.0393  2.3784  2.4495
0.2510  0.6606  0.6389  0.5774  0.6389  1.0000  1.0000  0.4518  0.9036  1.2574  0.4671
0.4387  0.7598  0.4204  0.7071  0.7953  1.0000  1.0000  0.3593  0.8409  1.7321  1.0000
1.7321  1.8612  1.8803  2.0000  1.5137  2.2134  2.7832  1.0000  2.7108  2.7108  2.5149
0.6148  0.3689  1.0746  0.8409  0.9622  1.1067  1.1892  0.3689  1.0000  2.1407  1.7602
0.3207  0.5373  0.5170  0.4041  0.4204  0.7953  0.5774  0.3689  0.4671  1.0000  0.6148
0.2510  0.3021  0.6223  0.5373  0.4082  2.1407  1.0000  0.3976  0.5681  1.6266  1.0000] % extended pairwise comparison matrix of benefit criteria
% consistency test
[n,n]=size(M) % get the dimension of M
[V,lambda]=eig(M) % returns diagonal matrix lambda of eigenvalues and matrix V whose columns are the corresponding right eigenvectors
lambda_max=max(max(lambda)) % find the maximum eigenvalue lambda_max
CI=(lambda_max-n)/(n-1) % compute the consistency index (CI) of M
RI=[0.00,0.00,0.58,0.90,1.12,1.24,1.32,1.41,1.45,1.49,1.51,1.48,1.56,1.57]
CR=CI/RI(n)
% calculating weights
[~,c]=find(lambda_max==lambda,1) % find the index of the principal vector (as it's stored in a diagonal matrix, we take the column index)
%The principal eigenvector is then V(:,c)
W1=V(:,c)./sum(V(:,c)) % normalization of principal eigenvector, to get the weight of the benefit model
CriBenefit=[1782.3  31463.9 0.0089  7.2 1   0   1870    250 370 4000    0
15500   4559    0.0067  61.2    0.73    500 14400   250 369 200 0
7833    56350   0.1984  84  0.8 0   3120    250 1000    300 1000
1000    280000  0.00000088  11.5    0.8 0   1345    250 832639  0.083   0
2918    124020  0.00000088  11.5    0.8 0   1345    250 1451    0.167   0
3500    8500    0.056   49.8    1   0   2850    250 137 50  0] % The benefit model's criterion-wise scoring matrix
%min-max normalization of the scoring matrix
for c=1:11
    minm=min(CriBenefit(:,c))
    maxm=max(CriBenefit(:,c))
    for r=1:6
        if(maxm-minm~=0)
        CriBenefit(r,c)=(CriBenefit(r,c)-minm)/(maxm-minm)
        end
        if(maxm-minm==0)
        CriBenefit(r,c)=1
        end
    end
end
%add weights to the final benefits score
for r=1:6
    BS(r)=0
    for c=1:11
        BS(r)=BS(r)+CriBenefit(r,c)*W1(c)
    end
end
%output: BS = [0.3364    0.4984    0.5220    0.3986    0.2714    0.3766]

% Part 2: cost model
M=[1.0000   0.7401  1.4142  2.7108  2.9130
1.3512  1.0000  2.0000  2.2795  2.5900
0.7071  0.5000  1.0000  1.4142  1.2779
0.3689  0.4387  0.7071  1.0000  0.3761
0.3433  0.3861  0.7825  2.6591  1.0000] %glass cost model criteria pairwise comparison matrix
% consistency test
[n,n]=size(M) % get the dimension of M
[V,lambda]=eig(M) % returns diagonal matrix lambda of eigenvalues and matrix V whose columns are the corresponding right eigenvectors
lambda_max=max(max(lambda)) % find the maximum eigenvalue lambda_max
CI=(lambda_max-n)/(n-1) % compute the consistency index (CI) of M
RI=[0.00,0.00,0.58,0.90,1.12,1.24,1.32,1.41,1.45,1.49,1.51,1.48,1.56,1.57]
CR=CI/RI(n)
% calculating weights
[~,c]=find(lambda_max==lambda,1) % find the index of the principal vector (as it's stored in a diagonal matrix, we take the column index)
%The principal eigenvector is then V(:,c)
W2=V(:,c)./sum(V(:,c)) % normalization of principal eigenvector
CriCost=[0.0013 1500000.0000    831.6151    484.0000    0.0000
0.0019  68000.0000  13500.0000  7857.0000   0.0000
0.2000  4853.0000   300.0000    174.6000    0.0000
0.0300  42000.0000  120.2749    70.0000 0.0000
0.0007  42000.0000  120.2749    70.0000 0.0000
0.0019  68000.0000  13500.0000  7857.0000   0.0000] % The criterion-wise scoring matrix of cost model
%min-max normalization
for c=1:5
    minm=min(CriCost(:,c))
    maxm=max(CriCost(:,c))
    for r=1:6
        if(maxm-minm~=0)
        CriCost(r,c)=(CriCost(r,c)-minm)/(maxm-minm)
        end
        if(maxm-minm==0)
        CriCost(r,c)=1
        end
    end
end
%adding weights to calculate costs score
for r=1:6
    CS(r)=0
    for c=1:5
        CS(r)=CS(r)+CriCost(r,c)*W2(c)
    end
end
%output: CS=[0.4746    0.4230    0.4275    0.1946    0.1538    0.4230]

%Part 3: final scoring
FS=BS-CS
%output: FS=[-0.138273314386487 0.0753313686218907  0.0945648183178742  0.203986921853510   0.117649745643192   -0.0464192734347949]
% our decision: recycle paper, cardboard, and rigid plastic
```